\documentclass[%
 reprint,
 amsmath,amssymb,
 aps,
prl,
]{revtex4-2}

\usepackage{graphicx}
\usepackage{dcolumn}
\usepackage{bm}
\usepackage[colorlinks=true, allcolors=blue]{hyperref}
\usepackage{mathrsfs}%
\usepackage[title]{appendix}%
\usepackage{xcolor}%
\usepackage{textcomp}%
\usepackage{booktabs}%
\usepackage{algorithm}%
\usepackage{algorithmicx}%
\usepackage{algpseudocode}%
\usepackage{listings}%
\usepackage{float}
\usepackage{siunitx}

\begin{document}

\title{Electron Acceleration in a Flying-Focus Laser Wakefield Accelerator}

\author{Aaron Liberman}
\altaffiliation{These authors contributed equally \\ Corresponding Emails: \\ aaronrafael.liberman@weizmann.ac.il,\\ victor.malka@weizmann.ac.il}
\author{Anton Golovanov}%
\altaffiliation{These authors contributed equally \\ Corresponding Emails: \\ aaronrafael.liberman@weizmann.ac.il,\\ victor.malka@weizmann.ac.il}
\author{Slava Smartsev}%
\altaffiliation{These authors contributed equally \\ Corresponding Emails: \\ aaronrafael.liberman@weizmann.ac.il,\\ victor.malka@weizmann.ac.il}
\author{Anda-Maria Talposi}%
\altaffiliation{These authors contributed equally \\ Corresponding Emails: \\ aaronrafael.liberman@weizmann.ac.il,\\ victor.malka@weizmann.ac.il}
\author{Sheroy Tata}%
\author{Victor Malka}%
\affiliation{%
 Department of Physics of Complex Systems, Weizmann Institute of Science, Rehovot 7610001, Israel
}%

\date{\today}

\begin{abstract}
Structured light pulses hold significant promise for their ability to overcome dephasing in laser-wakefield accelerators, that should facilitate applications in high-energy physics and XFEL. Numerical studies have shown that sculpting a pulse into a flying focus and using it to drive a wakefield can achieve dephasing-free acceleration of electrons, with gain in excess of 100\,GeV within reachable with existing laser facilities. This work reports on novel experiments using a flying-focus generated laser-wakefield accelerator to accelerate electrons to relativistic energies. The flying-focus pulse is achieved by sculpting the laser-pulse before focusing using spatio-temporal couplings and generating a quasi-Bessel beam with an axiparabola. This combination allows for the tuning of the propagation velocity of the wakefield, which, we demonstrate, has an impact on the maximum achievable electron energy. Optical and particle-in-cell simulations are used to support the data and to provide direct evidence of the partial mitigation of dephasing through this flying-focus scheme. These results are further elucidated in our companion letter \cite{Liberman_PRR_2026}.

\end{abstract}

\maketitle

\section{I. Introduction}

Laser-wakefield accelerators (LWFAs) \cite{Tajima_PRL_1979} have shown great promise in the pursuit of the miniaturization of the particle accelerator. LWFAs have gone from the initial demonstration of monoenergetic electron beams at energies of around 100 MeV \cite{Faure_Nature_2004,Geddes_Nature_2004,Mangles_Nature_2004} to the achievement of electron energies exceeding 10 GeV \cite{Picksley_PRL_2024, Rockafellow_PoP_2025}. In this time, significant research has been done in the use of LWFAs in various applications. These include the development of electron based radiation therapy \cite{Glinec_Med_Phys_2006}, high resolution medical imaging \cite{Malka_NaturePhysics_2008}, and the development of compact free-electron lasers \cite{Wang_Nature_2021,Labat_NaturePhotonics_2022,LaBerge_NaturePhotonics_2024}. Recently, work has begun on using LWFA-generated electron beams to study strong-field quantum electrodynamics \cite{Mirzaie_NaturePhotonics_2024,LUXE_EurPhysJ_2024}. For the continued growth of LWFA based applications, however, even higher electron energies must be achieved. A significant challenge that must be contended with is the dephasing limit. This is an energy limit imposed by the fact that the accelerated electrons outpace the wakefield, preventing further acceleration from taking place \cite{Joshi_Nature_1984,Esarey_ReviewOfModernPhysics_2009}. Overcoming the dephasing limit is the key to enabling LWFAs to rival and even exceed the most powerful conventional accelerators \cite{Caizergues_NaturePhotonics_2020,Palastro_PRL_2020,Shaw_PoP_2025}. In addition, beating dephasing could enable small-scale few-fs laser systems \cite{Faure_PPCF_2019} to achieve energies relevant for applications such as radiotherapy \cite{Caizergues_NaturePhotonics_2020}. 

Several approaches have been proposed and attempted in order to avoid the dephasing limit. These include adding a density upramp in the gas target---known as rephasing---which shrinks the wakefield size over the course of the acceleration, keeping the electrons in the accelerating phase for longer \cite{Sprangle_PRE_2001,Guillaume_PRL_2015,Gustafsson_ScientificReports_2024}. Another proposed technique is the use of multi-staged LWFAs \cite{Leemans_PhysicsToday_2009}. This would allow electrons to exit an LWFA prior to dephasing and then be injected into the accelerating phase of a new LWFA. The method which has, to date, yielded the most energetic electrons has been to decrease the plasma density in the LWFA, thereby minimizing the difference between the propagation velocity of the wakefield and the velocity of the electrons \cite{Picksley_PRL_2024,Leemans_NaturePhysics_2006,Gonsalves_PRL_2019,Miao_PRX_2022}. Through this technique, researchers have demonstrated the ability to accelerate electrons with energies exceeding 10 GeV \cite{Picksley_PRL_2024, Rockafellow_PoP_2025}. The challenge with this method, however, is how it scales to achieve ever greater electron energies. Lowering the plasma density also lowers the acceleration gradient \cite{Tajima_PRL_1979}, thus requiring greater acceleration lengths to achieve the same energy. This creates an increasingly challenging problem of laser diffraction, requiring ever most sophisticated guiding methods \cite{Esarey_ReviewOfModernPhysics_2009,Leemans_NaturePhysics_2006}. The challenges grow nonlinearly, stunting the ability to make big jumps in electron energy. 

The flying-focus pulse \cite{Sainte-Marie_Optica_2017,Froula_NaturePhotonics_2018,Debus_PRX_2019,Caizergues_NaturePhotonics_2020,Palastro_PRL_2020}---a pulse with a focus extended over several Rayleigh lengths, with different rays focusing to different points along the optical axis---is a different paradigm for overcoming dephasing  \cite{Debus_PRX_2019,Caizergues_NaturePhotonics_2020,Palastro_PRL_2020}. It relies on the ability to tune the propagation velocity of the laser driver in the plasma and to match it to its velocity in vacuum. The flying focus has been theorized and demonstrated with several experimental setups, which include colliding two tilted laser pulses \cite{Debus_PRX_2019} and combining second order spectral phase with longitudinal chromatism \cite{Sainte-Marie_Optica_2017,Froula_NaturePhotonics_2018}, known as the ``chromatic flying-focus.'' These techniques, however, are experimentally challenging to realize for high intensity, broadband laser pulses.

\begin{figure*} [t!]
   \begin{center}
    \includegraphics[width=\linewidth]{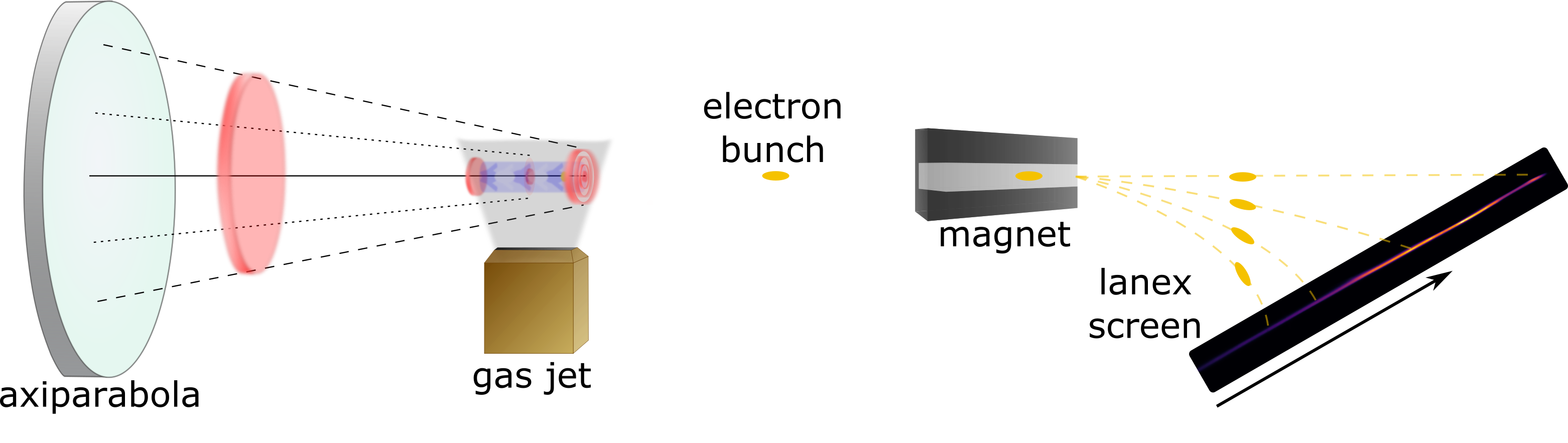}
    \end{center}
    \caption[]{ \label{fig:setup}  Schematic representation of the electron acceleration experiment. A laser pulse (red disk) is focused by the axiparabola (turquoise cylinder) onto a gas jet (gray emanating from gold jet), creating a wakefield (blue column) and accelerating an electron bunch (yellow dot). The laser pulse is also shown at focus over the gas jet demonstrating the development of the Bessel rings along the focal depth. After the gas jet, the electrons travel through a magnet, which gives an energy dependent trajectory to the electrons, which then impinge on a Lanex scintillator screen.  A sample Lanex image with an electron spectrum is shown. This figure also appears in the companion paper \cite{Liberman_PRR_2026}}
\end{figure*} 

A promising path towards a high-intensity-compatible realization of the flying-focus beam is based on spatio-temporal manipulation of the pulse in the near-field and focusing via a long-focal-depth optical element known as the axiparabola \cite{Smartsev_OpticsLetters_2019,Oubrerie_JoO_2022,Caizergues_NaturePhotonics_2020,Palastro_PRL_2020}, which generates a quasi-Bessel beam at focus. This combination allows the tuning of the timing of focusing of different annular segments of the beam to different locations along the optical axis, thereby allowing a tunable velocity flying-focus \cite{Caizergues_NaturePhotonics_2020,Oubrerie_JoO_2022,Ambat_OpticsExpress_2023,Palastro_PRL_2020,Liberman_OL_2024}. Since the proposal of this implementation scheme, significant efforts have been devoted towards understanding this new regime of LWFA and how it can be practically implemented \cite{Liberman_OL_2024,Ambat_OpticsExpress_2023,Pigeon_OE_2024,Geng_PoP_2022,Liberman_NatureCommunications_2025,Geng_ChinesePhys_2023,Abedi-Varaki_APA_2025,Ramsey_PRA_2023,Miller_ScientificReports_2023,Shaw_PoP_2025,Liberman_CLEO_2024}. Simulations have predicted that combining the axiparabola with spatio-temporal couplings allows for the acceleration of electrons with energies in excess of 100 GeV in meter-scale accelerators \cite{Caizergues_NaturePhotonics_2020,Palastro_PRL_2020,Shaw_PoP_2025}. Experimentally, the tunability of the on-axis propagation velocity has been demonstrated \cite{Liberman_OL_2024,Pigeon_OE_2024} and recent experiments have yielded insights into the structure of these flying-focus-based LWFAs \cite{Liberman_NatureCommunications_2025,liberman2025probingflyingfocuswakefields}. 

In this work, we introduce the first, to our knowledge, successful acceleration of electrons to relativistic energies with a flying-focus-based laser-wakefield. Acceleration with several wakefield velocities is demonstrated, achieved by manipulation of the pulse-front curvature (PFC)---a radially dependent spatio-temporal coupling---of the beam. As we show, the maximum cutoff energy of the electrons has a dependence on the wakefield velocity. Our results, both the electron spectra themselves as well as the energy dependence on the wakefield velocity, are reproduced in state-of-the-art particle-in-cell (PIC) simulations, and additional confirmation is brought in via an analytical model. The simulations yield additional insight into the acceleration mechanisms and show direct evidence of the partial mitigation of dephasing via the speeding up of the LWFA. The results show that the flying-focus-based wakefield can maintain the coherent structures that are required to accelerate electrons to relativistic energies, something that, until now, experiments have not succeeded in demonstrating. This proof-of-concept experiment is an important step towards the realization of truly dephasingless LWFAs.

\section{II. Experimental Methods}

\begin{figure*} [t!]
   \begin{center}
    \includegraphics[width=\linewidth]{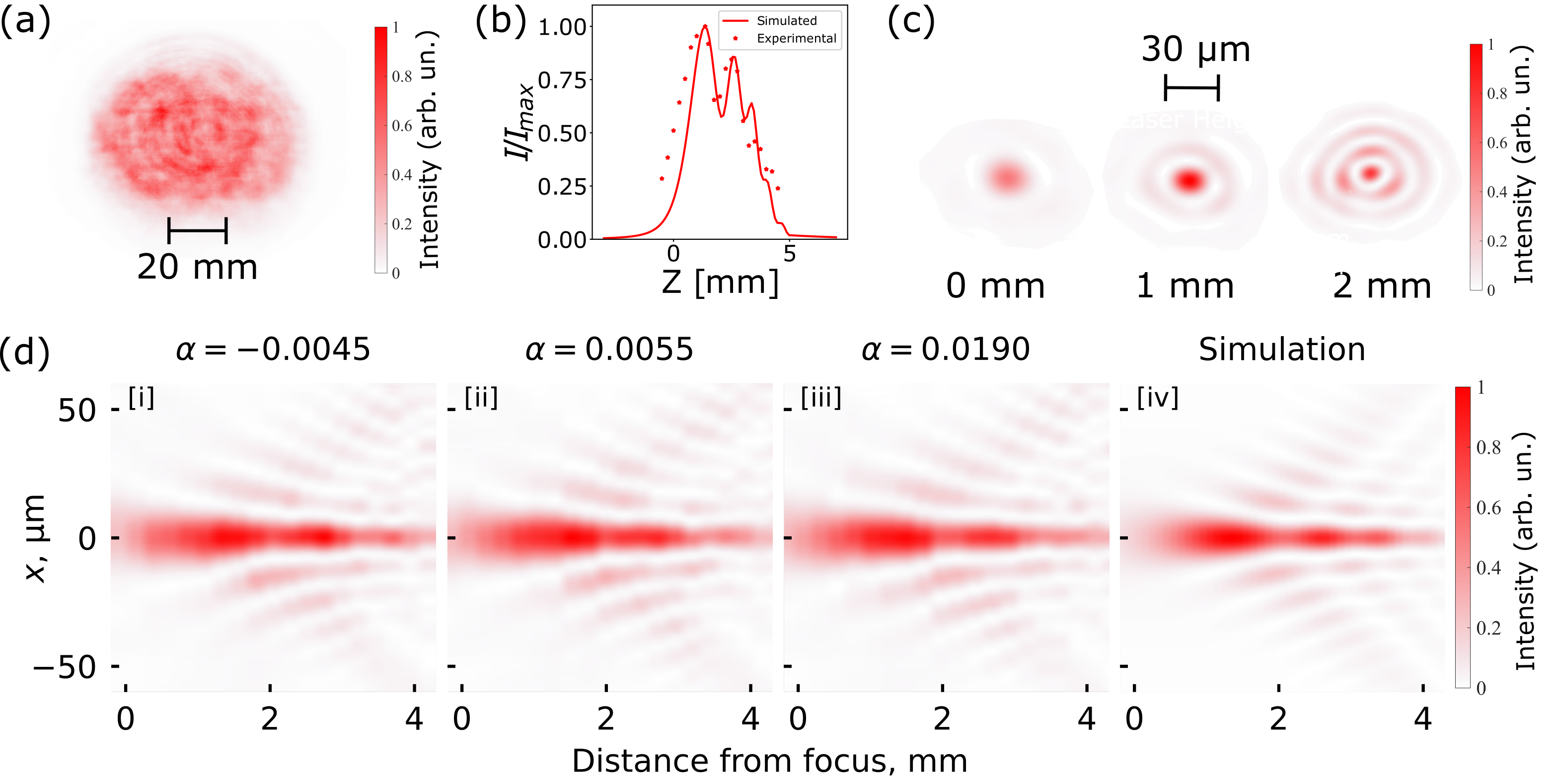}
    \end{center}
    \caption[]{ \label{fig:laser_params} (a) Typical 2D near-field profile of the laser. (b) Normalized focal spot intensity over the focal depth, in vacuum. Solid line shows simulated value while markers show experimentally measured points. (d) Selected 2D focal spots images at different points along the focal depth. (d) 2D (1 transverse direction $x$ and longitudinal direction $z$) fluence maps of the axiparabola focal-spot over the focal depth, generated from focal spot scans for (i) $\alpha = -0.0045$, (ii) $\alpha = 0.0055$, (iii) $\alpha = 0.0190$, and (iv) simulated (with $\alpha=0$) cases. The maps were generating by combining together experimental 2D focal scans along the focal depth, for each case, and then taking a 2D slice of the combined data. Figure 1 (b) and (c) appeared in the companion paper \cite{Liberman_PRR_2026}.}
\end{figure*} 

\subsection{Overview of Experimental Setup}

Figure \ref{fig:setup} provides a schematic of the experimental setup used. A pulse of the Weizmann Institute of Science's HIGGINS 100 TW laser system \cite{Kroupp_MRE_2022} is shown being focused by an axiparabola onto a slit nozzle, generating a flying-focus wakefield. The evolution of the axiparabola focal spot over the focal depth is illustrated. The resultant electron bunch then passes through a magnet and lanex spectrometer. A sample lanex image of electrons is shown in the figure.

\subsection{Laser System}

For the experiment, the Weizmann Institute of Science's HIGGINS $2\times 100$ TW laser system \cite{Kroupp_MRE_2022} was used. This system is a double chirped pulse amplification \cite{Strickland_OpticsComm_1985} based Ti:Sapphire laser that, for this experiment, yielded \SI{1.5}{J}, \SI{27}{fs} laser pulses with a \SI{30}{nm} bandwidth and a central frequency of \SI{800}{nm}. In the near-field, the laser has a diameter of around \SI{50}{mm} and a quasi-top-hat profile. Figure \ref{fig:laser_params} (a) shows a typical near-field profile of the beam. 

The laser was focused by an axiparabola designed so that its focal depth $\delta$ as a function of radius $r$ follows the form, $ f(r) = f_0 + \delta (r/R)^2$ \cite{Oubrerie_JoO_2022}, where $R$ is the full aperture of the beam. The axiparabola had an off-axis angle of 10 degrees, a focal depth of \SI{5}{mm} and a nominal focal length, $f_0$, of \SI{480}{mm} (f/9.6). Figure \ref{fig:laser_params} (b)  compares the measured (red markers) and simulated (solid red line) normalized intensity  of the axiparabola focused pulse over the focal depth. Figure \ref{fig:laser_params} (c) provides three measured 2D focal spot images at the start of the focal line, \SI{1}{mm} in, and \SI{2}{mm} in. Together, they show the development of the Bessel-ring structure. 

 Prior to entering the grating compressor, the laser went through a beam expansion telescope stage which brought the laser to its final diameter. The use of refractive beam expanders in the laser cause an inherent PFC. A specially designed doublet lens was inserted into this telescope, which was designed to modify the PFC of the beam \cite{Kabacinski_2021,Smartsev_JoO_2022, Liberman_OL_2024}. The impact of the doublet depends on the size of the incident beam, with a larger beam causing a greater modification of the PFC \cite{Kabacinski_2021,Smartsev_JoO_2022, Liberman_OL_2024}. When placed close to the first lens in the telescope, the doublet causes a partial suppression of the PFC inherent in the beam; when placed in the middle of the telescope, the doublet fully suppresses the PFC; and when placed near the second lens, the PFC is inverted \cite{Liberman_OL_2024}. The doublet was designed to modify the PFC while introducing negligible aberrations in the mean. Moving the doublet causes a small change in the focusing term of the beam, causing a \SI{400}{\um} focal shift between the most negative and most positive PFCs. The focal shift is predicted by Ansys Zemax OpticStudio simulations and was confirmed in experimental measurements. As the doublet can introduce pulse-front tilt (PFT) if is not perfectly centered, at each of the doublet positions, the PFT was optimized. 

 \begin{figure*} [t!]
   \begin{center}
    \includegraphics[width=\linewidth]{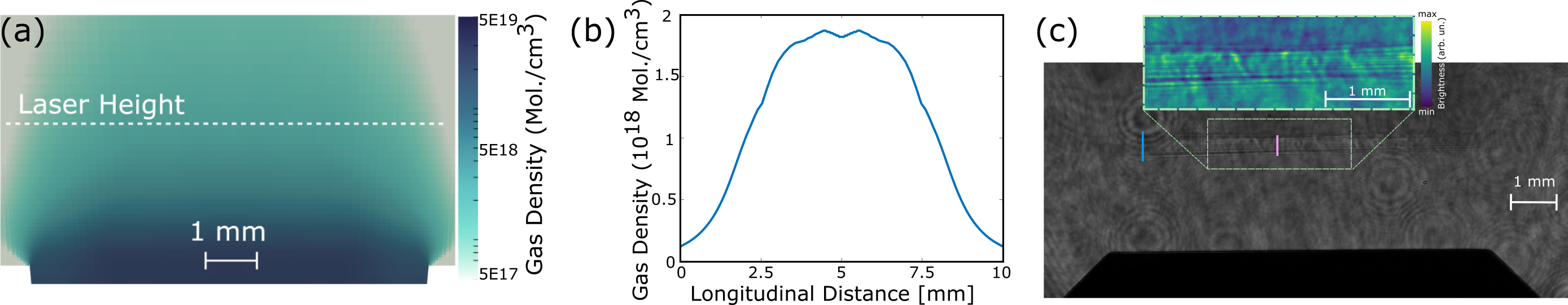}
    \end{center}
    \caption[]{ \label{fig:Nozzle_params} (a)  Ansys Fluent simulation of gas density from the nozzle. Dotted white line shows laser height. (b) 1D longitudinal density profile that corresponds to density at laser height in part (a). (c) Shadowgraphy image of the nozzle and plasma formed by the laser. The colored lines show the transverse size of the plasma at different points along the focal depth. The insert shows a closeup of the plasma in false color. Part (a) of the figure also appears in the companion paper \cite{Liberman_PRR_2026}}
\end{figure*} 

 The measurement of the PFC was done via far-field beamlet cross-correlation \cite{Liberman_OL_2024,Smartsev_JoO_2022}, a spatio-temporal metrology device that uses the interference of two beamlets in the far-field and is based on inverse Fourier transform spectroscopy\cite{Smartsev_JoO_2022}. More details on the PFC control doublet can be found in \cite{Smartsev_JoO_2022} and more details on the measurement can be found in \cite{Liberman_OL_2024}. 

To confirm that, besides the focal shift, the PFC shift was not introducing other significant distortions, at each PFC value a focal spot scan was performed. Figure \ref{fig:laser_params} (d) shows 2D fluence maps of the laser at focus for the PFC values – expressed as a spatio-spectral phase with a functional form $\alpha r^2 (\omega-\omega_0)$ – of (i) $\alpha = -0.0045$, (ii) $\alpha = 0.0055$, (iii) $\alpha = 0.0190$, and (iv) simulated (with $\alpha=0$) cases (in units of \si{fs/mm^2}). The maps are obtained by adding together the experimental focal scan and then making a 2D slice of the combined data. Along the Y-axis, one transverse direction is shown and along the X-axis the longitudinal direction is shown. A comparison of the cases shows that, besides for a longitudinal shift from the focal shift described above, there is no significant difference between the pulses indicating that changing the PFC value didn’t lead to a change in the transverse fluence, intensity distribution, and mode structure. The experimental images also closely match the simulated result for the parameters of the axiparabola, confirming that no unexpected modification to the phase front is introduced in the experiment. In the simulations (shown for $\alpha = 0$), the intensity map virtually does not depend on the value of $\alpha$ either, further reinforcing the argument that changing PFC does not affect the intensity distribution and the spot size.

\subsection{Gas Jet}

The gas jet used for the acceleration was an \SI{0.5}{mm} wide, \SI{7}{mm} long and \SI{36.8}{mm} high supersonic slit nozzle with a $\SI{500}{\um} \times \SI{550}{\um}$ throat. A mixture of 97\% helium and 3\% nitrogen was used, allowing for ionization injection of electrons into the wakefield. The gas profile generated by the nozzle was obtained through simulations, using the Ansys Fluent software, in which a full 3D geometric simulation of the nozzle was used. The nozzle was modeled with a mesh of around $10$ million elements, allowing for accurate solving of the flow equations and for proper evaluation of the thermal conductivity, specific heat, and viscosity parameters. The pressure at the outlet was initially set to $1$ Pa, while the inlet pressure was adapted to correspond to the experimental parameters, using $30$ bars of inlet pressure.  Figure \ref{fig:Nozzle_params} (a) shows a 2D cut of the output of the Fluent simulation where the white dotted line shows the laser height and the colorscale shows the gas density. Figure \ref{fig:Nozzle_params} (b) shows the longitudinal profile of the density output by the simulation, at the laser height of \SI{3}{mm} from the nozzle exit. Under the assumptions that the gas number density was the same for the helium-nitrogen mixture, helium was fully ionized, and nitrogen was partially ionized up to the fifth level, the estimated background electron plasma density was up to \SI{4e18}{cm^{-3}}. In order to ensure proper alignment, an unfocused probe beam, not shown in the setup figure, was passed through the wakefield. Figure \ref{fig:Nozzle_params} (c) shows a shadowgraphy image obtained from this probe beam, showing both the gas target and plasma at a height of $3$ mm above the target. The plasma width evolves over the course of the focal depth, with the radius at the location of the blue line being around \SI{340}{\um} while at the purple line it is around \SI{220}{\um}. The zoomed insert shows a detailed close up of the plasma in the shadowgraphy in false color. 

The simulation accuracy was validated by comparing interferometry measurements of the slit nozzle using argon gas to simulated flow of this nozzle using argon. Argon was used for the interferometer measurements due to the significantly higher phase shift that it imposes, allowing for more accurate measurement. The close match between the simulated and experimentally reconstructed density profiles in argon attests to the accurate performance of the simulation. 

In order to account for the focal shift of the laser for different PFC values, the gas jet was moved a corresponding distance, ensuring that for each PFC the laser was focused at the same focal depth.

\subsection{Spectrometer}

Electron bunches accelerated in the LWFA then passed through a 20-cm-long 1 T dipole magnet. The magnet imposed a curve to the trajectory of the electrons in the bunch, with the angle of the shift depending on the energy of the electron. Thus, when the electrons were allowed to propagate after the magnet, they spread in the $x$-axis direction, spatially separating out according to energy. The electrons then impinged onto a Lanex scintillating screen and the resulting scintillation was captured by a Hamamatsu ORCA-FLASH4.9 digital CMOS camera.

The relationship between the $x$-axis position on the lanex and the electron energy was calculated through simulations of the electron trajectories. The field of the magnet was mapped out experimentally and, along with the geometry of the experimental setup, was input into simulations to calculate the electron trajectories. Meanwhile, the $y$-axis behavior of the electrons on the Lanex yields information about the divergence of the electron bunch. The conversion from pixel brightness to absolute charge was done by calibrating the Lanex with a radioactive tritium source. 

\section{III. Simulation Methods}

Simulating electron acceleration with a flying-focus wakefield requires a multipart simulation. To begin with, the axiparabola-focused beam was initialized, reflected from the axiparabola, and propagated to focus via Axiprop \cite{Andriyash_Axiprop,Oubrerie_JoO_2022}, a code developed specifically for such long-focal-depth beams. The beam, prior to reflection from the axiparabola, was, transversely, a 16th order super-Gaussian with an intensity FWHM diameter of \SI{48}{mm} and, an energy of 1.5 J and, and a Gaussian temporal profile of 30 fs duration (intensity FWHM). Figure \ref{fig:setup} (b) shows the simulated intensity profile on-axis, compared to the measured intensity profile. In order to simulate the PFC, an additional phase was applied to the beam prior to reflection from the axiparabola. Since the Axiprop solver is axisymmetric and to accommodate for the angular mode decomposition used later on in the PIC code, an on-axis axiparabola, which otherwise had corresponding parameters to those found in the experiment, was implemented in the code. 

The beam that was output by Axiprop was then saved in the LASY format \cite{thevenet_lasy} and was used to initialize a PIC simulation using the quasi-3D spectral code FBPIC \cite{Lehe_ComPhysCom_2016}. The simulation was run with the laser polarized linearly in the $x$ direction (p-polarization). To be certain that the simulation captured all of the laser interaction, a simulation using two azimuthal modes and a box with the dimensions of \SI{558}{\um} in the $r$ direction (transverse) and \SI{150}{\um} in the $z$ direction (longitudinal) was employed. As figure \ref{fig:Nozzle_params} (c) shows, this box dimension was sufficient to capture all the relevant laser interaction. The grid resolution was $dz = \SI{0.04}{\um}$ and $dr = \SI{0.24}{\um}$ and checks were made to ensure that the resolution was fine enough that it did not play a role in the physical results. The simulation was run in a Lorentz-boosted frame, with a boost factor $\gamma = 3$ \cite{Lehe_2016_PRE_94_53305}, in order to decrease the computational load. The axiparabola was initialized in the FBPIC code such that the start of the axiparabola focal line (in vacuum) was \SI{4}{\mm} into the gas target. The $I_\mathrm{max}$ of the simulation was \SI{1.32e19}{W/cm^2}.

The gas profile input into the simulation was taken from the Ansys Fluent simulation, where the 2D profile can be seen in figure \ref{fig:Nozzle_params} (a) and the density profile along the $z$ axis can be seen in figure \ref{fig:Nozzle_params} (b). The maximal density of background electrons was around \SI{4.2e18}{cm^{-3}}. A mixture of helium and nitrogen was employed and to match the experimental parameters the macroparticle was arranged such that there was a 3\% molecular share of $N_2$. Since ionization effects have been shown to significantly impact the wakefield structure in flying-focus beams \cite{Miller_ScientificReports_2023,liberman2025probingflyingfocuswakefields}, the gas was neutral at initialization. Within a radius of \SI{30}{\um}, 32 helium and 32 nitrogen neutral atomic particles were initialized in each 2D cell (2 in $r$, 2 in $z$, 8 in $\theta$ directions, respectively) while outside this radius, 8 helium and 8 nitrogen atomic particles (1 in $r$, 1 in $z$, 8 in $\theta$) were initialized in each 2D cell.

To convert the FPBIC generated electrons into a Lanex image, the electrons were computationally advanced through a model of the magnet and Lanex spectrometer, where the parameters of the magnet, the Lanex, and the overall geometry matched the experimental setup. Checks were done to ensure that the simulated calibration matched with the experimental one.

\section{IV. Results}

\begin{figure*} [t!]
   \begin{center}
    \includegraphics[width=\linewidth]{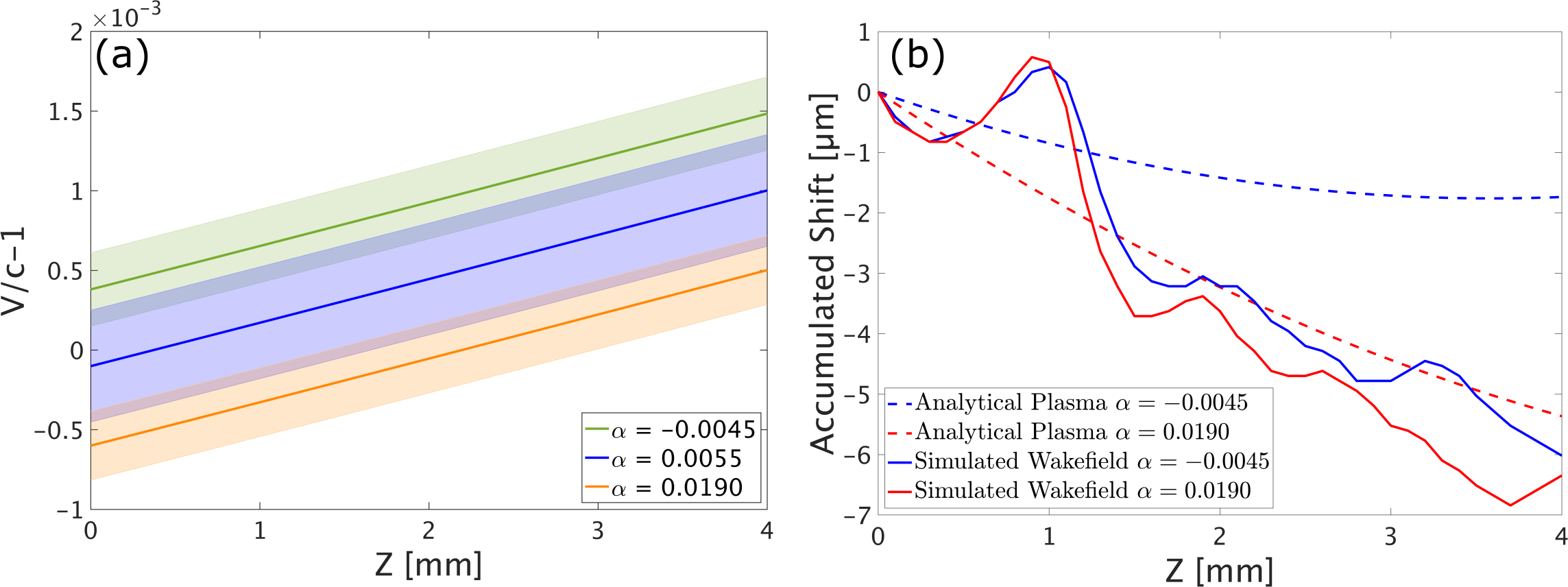}
    \end{center}
    \caption[]{ \label{fig:Velocity_Figure} (a) Measured velocity  of intensity peak propagation in vacuum along the optical axis for the axiparabola focused beam. Shown for the $\alpha = -0.0045$ (green), $\alpha = 0.0055$ (blue), and $\alpha = 0.0190$ (orange) cases. The shaded area corresponds to the measurement error. (b) Accumulated shift, compared to luminal propagation of the in-plasma analytical solution of the laser pulse for $\alpha=-0.0045$ (blue, dashed) and $\alpha = 0.0190$ (red, dashed) and the zero-point of the simulated wakefield for $\alpha=-0.0045$ (blue, solid) and $\alpha = 0.0190$ (red, solidy). Part (a) of the figure also appears in the companion paper \cite{Liberman_PRR_2026}}
\end{figure*} 

\subsection{Velocity of Intensity Peak}

For the axiparabola used in the experiment, when combined with an applied PFC in the near-field, the analytical expression of the velocity of propagation of the peak of intensity as a function of the focal depth is \cite{Liberman_OL_2024}: 

\footnotesize
\begin{equation}
    \frac{v_z}{c} = 1 - \frac{c\alpha R^2}{\delta}+\bigg(\frac{R^2}{2\delta f_0^2} - \frac{c\alpha R^4}{\delta^2 f_0^2}\bigg)z-\bigg(\bigg[\frac{R^2}{2\delta f_0^2} - \frac{c\alpha R^4}{\delta^2 f_0^2}\bigg]\frac{2}{f_0} -  \frac{c\alpha R^6}{2\delta^3 f_0^4}\bigg)z^2 \label{eq:full_expression}
\end{equation} 
\normalsize 

\noindent where $c$ is the speed of light, $\alpha$ is the PFC value, $R$ is the total radius of the pulse, $\delta$ is the total focal depth, $f_0$ is the nominal focal length of the axiparabola, and $z$ is the point along the focal depth. 

The velocity of the intensity peak was measured for the three PFC values used.  Figure \ref{fig:Velocity_Figure} (a) shows the measured deviation from luminal propagation velocity of the intensity peak along the focal depth, $v_z/c-1$. The figure shows the results for the $\alpha = -0.0045$ (green), $\alpha = 0.0055$ (blue), and $\alpha = 0.0190$ (orange) cases. The measurement was based on a modified version of far-field beamlet cross-correlation \cite{Smartsev_JoO_2022}, a spatio-temporal measurement technique based on far-field interferometry and inverse-Fourier-transform spectroscopy. Additional information about the methodology of the velocity measurements can be found in Ref. \cite{Liberman_OL_2024}. As can be seen in figure \ref{fig:Velocity_Figure} (a), the negative PFC corresponds to an intensity peak that propagates faster (superluminal in vacuum), while the positive PFC corresponds to an intensity peak propagating more slowly (sub-luminal in vacuum). 

The propagation velocity of the wakefield in the plasma, however, cannot be neatly correlated to the vacuum velocity. To characterize the magnitude of the shift, figure \ref{fig:Velocity_Figure} (b) shows a comparison of the accumulated shift, in microns, between the $\alpha = -0.0045$ and $\alpha = 0.0190$ cases for pulses focused by the axiparabola, along the focal depth. The shift is shown relative to traveling at the speed of light in vacuum. The dashed blue line shows the analytical solution, in plasma, for the $\alpha = -0.0045$ pulse. This solution was obtained by taking the analytical solution in vacuum according to Eq.~\eqref{eq:full_expression} and modifying it by the laser group velocity in plasma to account for the change in refractive index. The dashed red line shows the in-plasma analytical solution for the $\alpha = 0.0190$ case. 

The solid blue and red lines in figure \ref{fig:Velocity_Figure} (b) show the accumulated shift for the zero-point of the wakefield in the $\alpha = -0.0045$ and $\alpha = 0.0190$ cases, respectively. As can be seen, the accumulated shift for the wakefields themselves differs greatly from the analytical propagation of the laser pulse. The plasma plays a complex role in changing the relative propagation velocity between the two wakefields. There are many different causes for this divergence in behavior. Among these is the different distances the different annular segments of the beam travel in gas; the different moments that the annular segments ionize and, therefore, the uneven amount of plasma that they travel in; and the evolving size of the wakefield over the focal depth. 

Earlier work \cite{Liberman_NatureCommunications_2025} showed that, in spite of this, there is a velocity shift in the wakefields, with the wakefield generated with a beam that has negative PFC propagating faster than that which was a positive PFC. Crucially, a direct comparison of the propagation velocities of the wakefield here yields the same conclusion. Throughout the acceleration process, the $\alpha = -0.0045$ wakefield remains positively shifted when compared to the $\alpha = 0.0190$ wakefield, meaning that the wakefield generated by the $\alpha = -0.0045$ pulse is indeed traveling at a higher velocity.

\begin{figure*} [t!]
   \begin{center}
    \includegraphics[width=\linewidth]{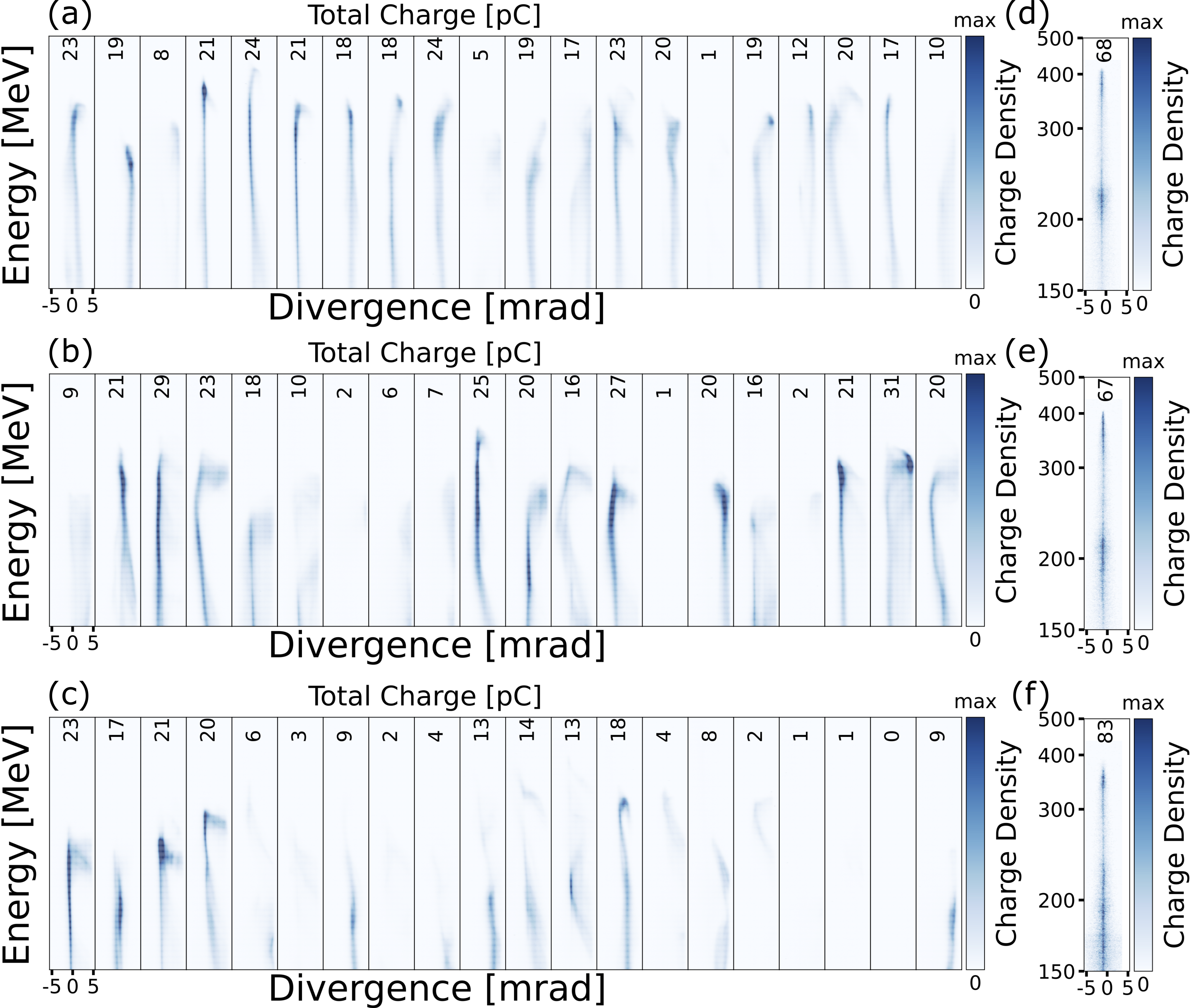}
    \end{center}
    \caption[]{ \label{fig:Data_Fig} (a--c) 20 Lanex images each for the $\alpha = -0.0045$ (b), $\alpha = 0.0055$ (c), and $\alpha = 0.0190$ (d) cases. The $x$-axis shows the divergence, the $y$-axis gives energy resolution, and the  colorbar provides charge density. Total charge above 150 MeV is provided for each shot. (e--f) Corresponding simulated Lanex images for the $\alpha = -0.0045$ (d), $\alpha = 0.0055$ (e), $\alpha = 0.0190$ (f) cases.}
\end{figure*} 

\subsection{Lanex Images}

For each PFC value, $\alpha$, of the beam, 20 shots were taken and the accelerated electrons were passed through the magnet-Lanex spectrometer. Figure \ref{fig:Data_Fig} shows the Lanex images after the spectrometer. The images are shown in an unprocessed form except for background subtraction. The Lanex images give information about the spectrum and divergence of the accelerated electrons. Since the length of the electron trajectories between the LWFA and the Lanex differ somewhat for different energies, the divergence shown in figure \ref{fig:Data_Fig} is the divergence for the center-point of the Lanex. The maximal error that this causes at the extremes of the Lanex is around 10\% of the given divergence values. The colorbar corresponds to the normalized charge density. The numbers at the top provide the total charge, in pico-Coulombs, that is above $150$ MeV. The $y$-axis extent of the Lanex has been cropped in order to focus on the area of interest, that found between $150-500$ MeV. 

An additional Lanex screen was placed before the entrance to the magnet prior to the spreading out of the electron bunch. By knowing the distance between this Lanex and the LWFA source, the spatial extent and the spatial jitter of the scintillation of the Lanex gives information about the pointing fluctuations and divergence of the beam. Pointing fluctuations in the horizontal direction were found to be around 5.5 mrad (RMS) while the divergence was around 1.5 mrad (RMS). At the cutoff energy of the electrons, the energy uncertainty is around 12 MeV.  

\begin{figure*} [t!]
   \begin{center}
    \includegraphics[width=\linewidth]{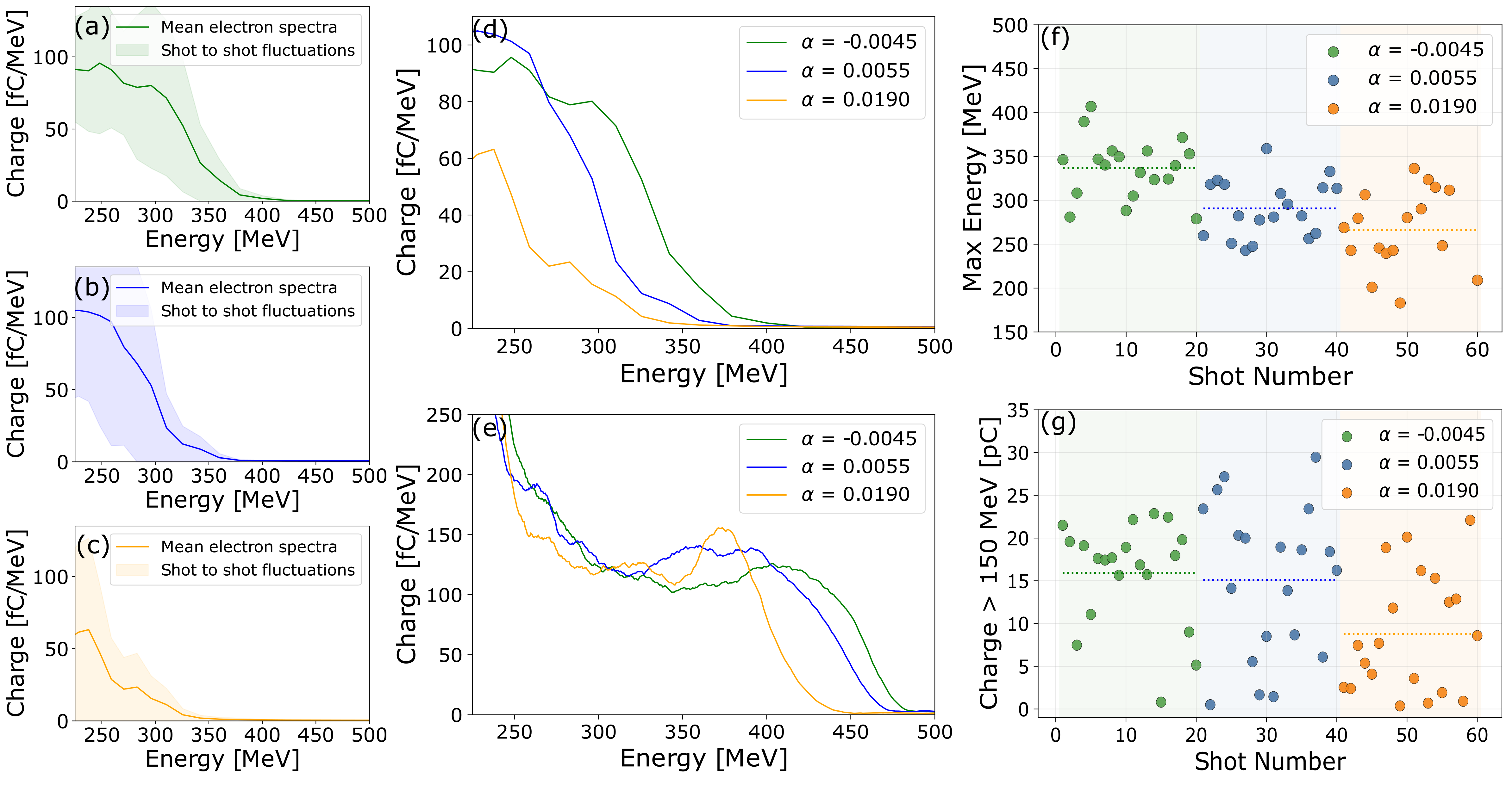}
    \end{center}
    \caption[]{ \label{fig:fig3} (a--c) Electron spectra above $225$ MeV, averaged over 20 shots, for the $\alpha = -0.0045$ (b, green), $\alpha = 0.0055$ (c, blue), and $\alpha = 0.0190$ (d, orange) cases. The energy is shown on the x-axis and the charge density on the y-axis. The shaded area is the RMS shot-to-shot fluctuation. (d) Comparison of the three averaged spectra for the three cases. (e) Comparison of the spectra obtained from the PIC simulation for the three cases. (f) Plot of the maximum energy fluctuations for each of the three cases. Dotted horizontal line gives average maximum energy for each of the cases. (g) Plot of the charge (above $150$ MeV) fluctuations for each of the three cases. Dotted horizontal line gives average charge for each of the cases. This figure also appears in our companion paper \cite{Liberman_PRR_2026} }
\end{figure*} 

Figure \ref{fig:Data_Fig} (a) shows 20 Lanex images for the $\alpha = -0.0045$ case.  As can be seen, the Lanex images show consistent acceleration of electrons to energies of hundreds of MeV with tens of pico-Coulomb charges.  Figure \ref{fig:Data_Fig} (b) shows the 20 shots for the $\alpha = 0.0055$ case and figure \ref{fig:Data_Fig} (c) shows the images for the $\alpha = 0.0190$ case. Comparing the Lanex images for the different velocities shows clearly that the maximal achievable electron energy has a dependence on the PFC value, and, therefore on the wakefield velocity. While the fastest wakefield ($\alpha = -0.0045$) achieves a maximum electron cutoff energy of around $400$ MeV, the slowest wakefield ($\alpha = 0.0190$) does not exceed $350$ MeV with the middle velocity wakefield ($\alpha = 0.0055$) having a cutoff in between. As was shown in figure \ref{fig:laser_params} (d), besides the PFC, the pulses are practically identical. Therefore, there is significant evidence that the maximum cutoff energy shift is directly caused by the impact of the change in wakefield velocity, suggesting that the speeding of the wakefield is partially mitigating dephasing effects.

It is noteworthy that in addition to the difference in maximum electron energy, the faster wakefield also appears to provide a more shot-to-shot stable acceleration of electrons than the slower wakefield. In the $\alpha = -0.0045$ case, two of the 20 shots would be classified as failing to accelerate significant charge of high energy electrons. In comparison, in the $\alpha = 0.0190$ case 4-5 of the 20 shots failed, while in the middle case 3 shots failed. 

To confirm our understanding of the results, the experimental results were compared to PIC simulations. Figures \ref{fig:Data_Fig} (d),(e), and (f), contain the PIC simulated Lanex images for the $\alpha = -0.0045$, $\alpha = 0.0055$, and $\alpha = 0.0190$ cases, respectively. As can be seen, the PIC simulations maintain the same velocity dependence of the maximum cutoff energy that is seen in the experimental Lanex images, with the simulated difference between the fastest and slowest cases being almost $50$ MeV, nearly the same energy difference that is seen in the experiment. It is important to note that the simulation contains a somewhat higher overall cutoff energy, around $485$ MeV for the $\alpha = -0.0045$ and $440$ MeV for the $\alpha = 0.0190$ case. In addition, the accelerated electron charge in the simulation is significantly higher than what is seen in the experiment. This is explained by the ideal, aberration-free axiparabola used in the simulation which did not take into account the impact from the aberrations that can be seen in the laser profile in figure \ref{fig:laser_params} (c) as well as the ideal Gaussian spectrum and temporal profile of the pulse. Moreover, the gas profile used in simulations, though taken from the Fluent results, does not account for possible imperfections of the nozzle which may cause some density fluctuations. Besides these minor inconsistencies, the remarkably close replication of the dependence of the maximum electron energy on the wakefield velocity gives significant evidence of the veracity of this effect, demonstrating the impact that increasing the wakefield velocity can have on the electron acceleration process.

\subsection{Statistical Analysis}

The trend seen in the raw Lanex images is born out even more clearly when a statistical analysis of the data is performed. Figure \ref{fig:fig3} (a--c) show graphs of electron energy versus charge density for the shots shown in figure \ref{fig:Data_Fig} (a--c), for the $\alpha = -0.0045$, $\alpha = 0.0055$, and $\alpha = 0.0190$ cases, respectively. The electron spectra are plotted above $225$ MeV to emphasize the cutoff energy and they are binned in order to account for the uncertainty in the energy due to beam pointing and divergence. The solid line shows the averaged spectra while the shaded region gives the RMS shot-to-shot fluctuation of the charge in each energy bin. 

Figure \ref{fig:fig3} (d) provides an overlap plot of the three averaged spectra, allowing for comparison. The figure clearly shows the significant dependence of the cutoff energy on the wakefield velocity, illustrating the $50$ MeV difference between the fastest wakefield ($\alpha = -0.0045$, green) and the slowest wakefield ($\alpha = 0.0190$, orange), with the intermediate velocity wakefield ($\alpha = 0.0055$, blue) falling in the middle. The experimental averaged spectra can be compared to the simulated spectra for the three wakefield velocities, shown in figure \ref{fig:fig3} (e). As expected, the simulated spectra closely follow the same dependence of the maximal energy on the velocity. As with the Lanex images, the cutoff energy of the simulations is somewhat higher and the charge is significantly higher when compared to the experimental results. However, the energy gap between the wakefield velocities is almost perfectly replicated. 

Yet another way of seeing this effect is by looking at the distribution of the maximal electron energies in each shot for the different PFC cases, as is shown in figure \ref{fig:fig3} (f). Each dot represents the cutoff energy for a particular shot and the dotted horizontal lines provide the averaged cutoff energy at each PFC. In addition to the clear visual discrepancy between the averaged cutoff energies, with the faster wakefield ($\alpha = -0.0045$, green) having a higher cutoff than the slower wakefields, the distributions differ in a statistically significant way, with a p-value of $0.0004$, giving a significance of above $3.5\sigma$ between the $\alpha = -0.0045$ and the $\alpha = 0.0055$ cases and significantly higher when comparing the $\alpha = -0.0045$ and the $\alpha = 0.0190$ cases. 

Notably, the charge above $150$ MeV does not behave in the same way as the energy, as if shown in figure \ref{fig:fig3} (g). The figure shows the charge for each shot, with the dotted horizontal lines showing the average charge for each PFC. As can be seen, while the charge fluctuates significantly shot-to-shot, between the $\alpha = -0.0045$ and the $\alpha = 0.0055$ cases there is no statistically significant shift in the average charge. The presence of the energy shift between them, therefore, cannot be explained away by beam loading effects, strengthening the evidence for it being directly caused by the change in wakefield velocity. 

\begin{figure}[t!]
    \centering
\includegraphics[width=\linewidth]{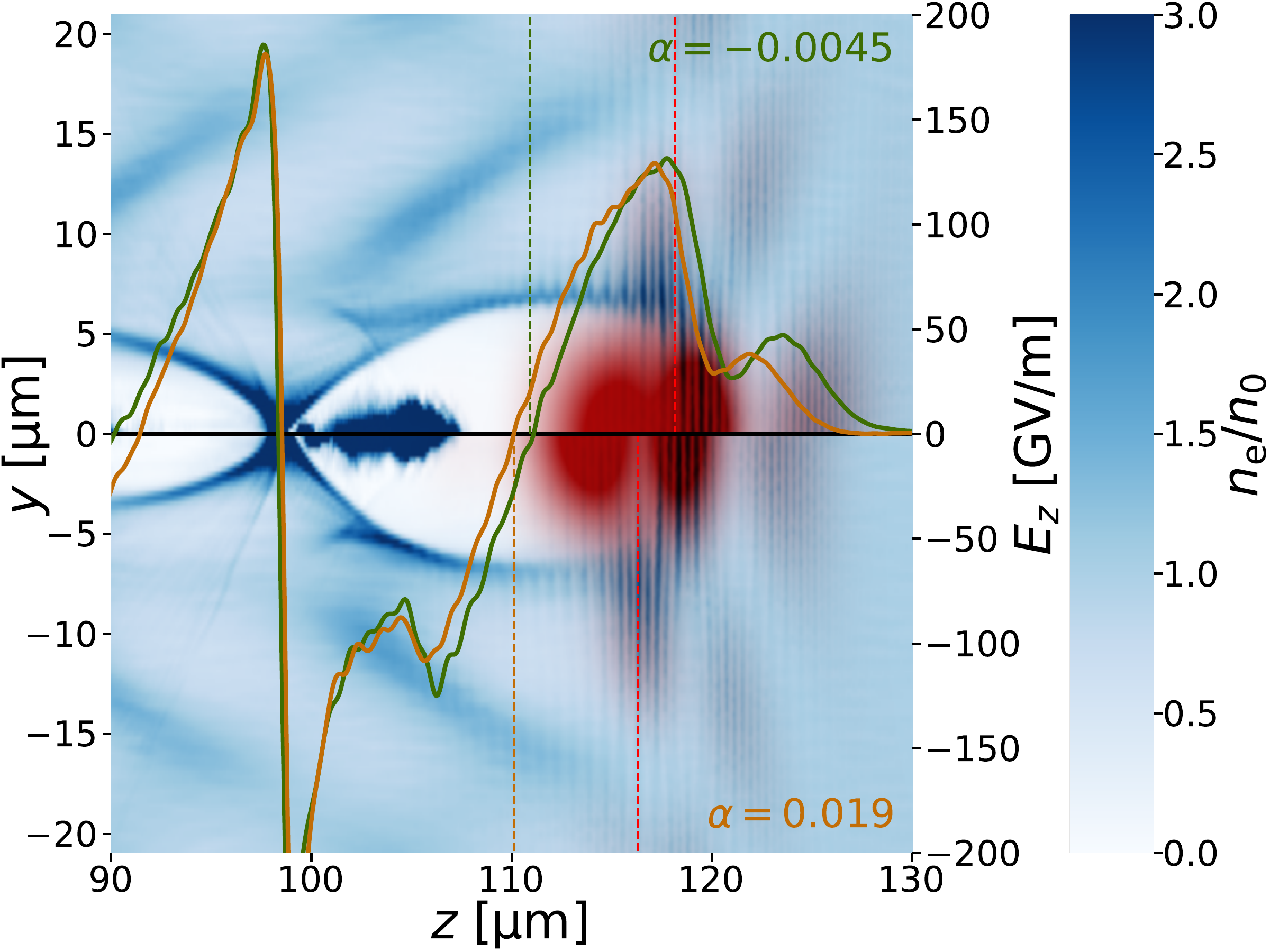}
    \caption{Simulated wakefield for $\alpha = -0.0045$ (top) and $\alpha = 0.0190$ (bottom) cases at the propagation distance of \SI{6}{mm} (\SI{2}{mm} after the beginning of the focal line). Blue colorbar shows relative electron density distribution $n_e/n_0$ and red color shows the intensity of the axiparabola laser field. Line plots show the longitudinal electric field $E_z$ for the $\alpha = -0.0045$ (green) and $\alpha = 0.0190$ (orange) cases. This figure was adapted from the companion paper \cite{Liberman_PRR_2026}.}
    \label{fig:wake}
\end{figure}

\subsection{Evidence for Dephasing Mitigation}

Getting a more direct picture of the wakefield dynamics requires looking at a comparison of the wakefields themselves, for different PFCs. Figure \ref{fig:wake} compares the PIC simulated wakefields for the $\alpha = -0.0045$ (top) and $\alpha = 0.0190$ (bottom) cases. The white-blue colorbar is $n_e/n_0$, the electron density distribution normalized by the local electron plasma density assuming full ionization of the helium and ionization of 5/7 electrons of the nitrogen. The red colorbar, meanwhile, shows the intensity of the laser field. The solid lines display $E_z$, the longitudinal acceleration field, for the two cases, $\alpha = -0.0045$ (green) and $\alpha = 0.0190$ (orange). The dotted vertical lines show the COM of the laser driver (red, top is $\alpha = -0.0045$ and bottom is $\alpha = 0.0190$), the zero point $E_z$ for the $\alpha = -0.0045$ case (green), and the zero point $E_z$ for the $\alpha = 0.0190$ case (orange).

The plot provides direct confirmation of the wakefields traveling at different velocities, as can be seen by the longitudinal delay of over a micron between the $\alpha = 0.0190$ wakefield and the $\alpha = -0.0045$ wakefield. It is notable that the electrons from the $\alpha = 0.0190$ case are somewhat closer to the zero point of $E_z$ than those in the $\alpha = -0.0045$ case, providing direct evidence that there is a partial mitigation of dephasing that occurs due to the difference in propagation velocity between the two wakefields.

\section{V. Basic Analytical Model} 

To further reinforce the veracity of the impact of the wakefield velocity on the electron cutoff energy, a simple analytical model was derived. 

We assume that there is a linear relationship, in the bubble regime, between the longitudinal electric field, $E_z$, and the coordinate $\zeta = z-z_0(t)$, where $z_0$ is the location of the bubble center  \cite{Lu_PhysRevSTAB_2007}. $E_z$ has the functional form: 
\begin{equation}
    E_z(\zeta) =  \frac{m\omega_p^2}{2e}\zeta = \frac{e n_e}{2\epsilon_0} \zeta,
\end{equation} 
where $\epsilon_0$, is the vacuum permittivity and $e$, $m$, and $\omega_p$ are the electron charge, the electron mass, and the plasma frequency, respectively.

The group velocity of the laser inside of plasma takes the form \cite{Esarey_ReviewOfModernPhysics_2009}: 
\begin{equation}
\label{eq:velocity2}
    \frac{v_\mathrm{gr}}{c} = \bigg[1-\frac{\omega_p^2}{\omega_0^2}\bigg]^{0.5} = \bigg[1-\frac{n_e}{n_c}\bigg]^{0.5} 
\end{equation}
where $\omega_0$ is the laser frequency. Plugging in the plateau density from figure \ref{fig:Nozzle_params} (b), \SI{4e18}{cm^{-3}}, the resulting group velocity is $0.9989c$. 

Applying the assumption that the electrons are ultrarelativistic compared to the wakefield ($dz/dt \approx c$) and that the wakefield's phase velocity is a simple sum of the laser group velocity and the velocity correction from the axiparabola-focused pulse with a particular PFC, $\Delta v(z)$ (as calculated in \cite{Liberman_OL_2024,Oubrerie_JoO_2022} and shown in figure \ref{fig:Velocity_Figure} (a)), the velocity takes the following form
\begin{equation}
    v_\mathrm{ph}(z) = v_\mathrm{gr} + \Delta v(z)
    \label{eq:velocity_def}
\end{equation}
Now, the coordinate is solved for as a function of time:
\begin{align}
    &\frac{d\zeta}{dt} = c-v_\mathrm{ph}(ct),  \\
    &\zeta(t) = \zeta_\mathrm{min} + \int\limits_{0}^{ct} \left[1-\frac{v_\mathrm{ph}(z)}{c}\right] dz.  
\end{align}
where $\zeta_\mathrm{min} < 0$ is the coordinate at the location of electron injection. From here, the longitudinal momentum gain can be derived:
\begin{align}
&\frac{dp_z}{dt} = -eE_z \implies \frac{dp_z}{dz} = -\frac{e}{c}E_z = -\frac{e^2 n_e}{2\epsilon_0 c}\zeta,\\
&p_z = -\frac{e^2 n_e}{2\epsilon_0 c}\int\limits_{0}^{z}\bigg(  \zeta_\mathrm{min} + \int\limits_{0}^{z'} \left[ 1-\frac{v_\mathrm{ph}(z'')}{c}\right] dz''\bigg)dz' .
\end{align}
Now inserting the definition of the velocity from equation \ref{eq:velocity_def}, using the measured velocity profiles in vacuum from figure \ref{fig:Velocity_Figure} (a), and assuming that the in-plasma profiles correspond to the vacuum profiles minus the difference between the velocity of the laser driver in vacuum and plasma, as obtained in equation \ref{eq:velocity2}, the electron energy can be calculated.

The assumptions used in the model, specifically the assumptions about the wakefield propagation velocity, are not accurate, as clearly shown in \ref{fig:Velocity_Figure} (b), and ignore much complexity such as beam loading \cite{Rechatin_PRL_2009} and the laser evolution over the focal depth \cite{Miller_ScientificReports_2023,Liberman_NatureCommunications_2025}. Therefore, the model overestimates the maximum electron energy significantly. However, the model still contains some of the essential physics at play and when the model is used to calculate the cutoff energy for the cases of $\alpha = -0.0045$ and $\alpha = 0.0190$, it likewise predicts that the faster wakefield will have a higher electron cutoff energy, by around $70$ MeV for parameters similar to this experimental setup. Thus, while it cannot be used for accurate predictions of the electron energy, the alignment of the analytical model's predictions with those seen in the experimental data and in the PIC simulations lends further credence to the observed effect.

\section{VI. Discussion}

The demonstration of the ability of flying-focus wakefields to maintain the coherent structures necessary for accelerating electrons to relativistic energies represents an important breakthrough in the pursuit of dephasingless acceleration. Combining this axiparabola-focused wakefield with pulse-front curvature allowed for the exploration of the impact of wakefield velocity on the electron acceleration process. The observed dependence of the maximum achievable electron energy on the wakefield velocity serves as a further proof-of-concept that these flying-focus wakefields are a viable solution to the dephasing limitation of LWFA. The strength of this proof-of-concept was furthered by the direct evidence of the mitigation of dephasing provided by the simulated wakefield snapshot that was shown. The good correspondence between the experimentally observed energy dependence and that seen in the PIC simulations and with the results of the analytical model lends further credence to these assertions. 

The results, however, also demonstrate the challenges that must yet be overcome before dephasingless acceleration is truly realized. As was shown, the translation between the in-vacuum propagation velocity of the laser drive and the in-plasma propagation velocity of the wakefield is highly non-trivial. Fully optimizing the velocity of the laser driver will require further exploration to identify the correct parameter set and more sophisticated measurement and control over the spatio-temporal couplings in order to implement the parameters correctly. Most likely, fine-control over the wakefield velocity will require the ability to measure \textit{in-situ}. 

The technologies for spatio-temporal metrology and control are rapidly advancing, with the development of new single-shot spatio-temporal measurement devices \cite{Smartsev_OL_2024,Howard_NaturePhotonics_2025} and the development of meta-optics capable of fine control of light-structure and able to withstand high-intensity laser pulses \cite{Oliveira_Optica_2025}. Perhaps these technologies will enable the achievement of stable, flying-focus wakefields optimized for mitigating dephasing and allow for a manyfold gain in the achievable energy of LWFAs \cite{Palastro_PRL_2020,Caizergues_NaturePhotonics_2020}. If successful, these techniques can be extended to other realms, such as the laser-wakefield acceleration of ions \cite{Gong_PRL_2024} and other exotic wakefield configurations like LWFAs from helical beams for positron acceleration \cite{Vieira_PRL_2014}.

\section*{Acknowledgments}
The authors would like to thank Dr. Igor Andriyash for developing the axiprop code; Dr. Eitan Levine, Dr. Yinren Shou, Ivan Kargapolov and Salome Benracassa for constructive discussions; and Dr. Eyal Kroupp for helping manufacture the setup.

\section*{Data Availability Statement}
The raw data used in this work, as well as the the code, can be made available upon reasonable request to the authors. 

\section*{Conflict of Interest}
The authors declare no conflicts of interest.

\section*{Funding}
The research was supported by the Schwartz/Reisman Center for Intense Laser Physics, the Benoziyo Endowment Fund for the Advancement of Science, the Israel Science Foundation, la Fondation du Judaïsme Français, Minerva, Wolfson Foundation, the Schilling Foundation, R. Lapon, Dita and Yehuda Bronicki, and the Helmholtz Association.

\appendix

\section*{References}

\bibliographystyle{ieeetr}

\bibliography{Refs}

\end{document}